\documentclass[a4paper,10pt]{amsart}
\usepackage{english}
\usepackage{amsmath,enumerate, amsfonts, amssymb,amsthm}
\input xy
\xyoption{all}

\newtheorem{Def}{DEFINITION}[section]
\newtheorem{Theo}[Def]{THEOREM}
\newtheorem{Prop}[Def]{PROPOSITION}
\newtheorem{Rem}[Def]{REMARK}
\newtheorem{Lem}[Def]{LEMMA}
\newtheorem{Cor}[Def]{COROLLARY}

\DeclareMathOperator{\hess}{hess}

\title[Effective $3$-forms]{On symplectic classification of effective
  $3$-forms and Monge-Amp\`ere equations.}

\author{Bertrand BANOS}
\address{D\'epartement de Math\'ematiques\\ Universit\'e d'Angers\\ 2
  bd Lavoisier, 49045 Angers, France}
\email{banos@tonton.univ-angers.fr}

\begin{document}

\begin{abstract}
We complete the list of normal forms for effective
$3$-forms with constant coefficients with respect to the natural
action of symplectomorphisms in $\mathbb{R}^6$. We show that the
$3$-form which corresponds to the Special Lagrangian equation is among
the new members of the classification. The symplectic symmetry algebras
and their Cartan prolongations for these forms are computed and a local
classification theorem for the corresponding Monge-Amp\`ere
equations is proved.\\
\end{abstract}

\maketitle

A classical problem of the geometric theory of differential equations
is the problem of local equivalence: when do two differential equations
represent the same equation modulo local change of  dependent and
independent coordinates (a diffeomorphism on the corresponding jet
space)? We can consider this problem  to be a special case of the
general equivalence problem in differential geometry (see \cite{5},
chapter 7). This point of view enables us to recognize equivalent
structures, objets, etc.,  by means of a set of so-called
``scalar'' differential invariants.  Generally speaking, a
differential invariant of order at most $k$ is a smooth function on
the jet space $J^k$  invariant under the diffeomorphisms
generated by the local diffeomorphisms of the base manifold.

An important particular case of this problem is the question of
linearization: when is a differential equation
equivalent to a linear one? The Monge-Amp\`ere equations (MAE hereafter) provide a
natural class of nonlinear second order differential equations for
this problem. Sophus Lie posed in 1874 this problem of linearization for
2-dimensional MAE with respect to the (pseudo) group of contact
diffeomorphism $Ct$ and it was studied in the
classical works of G.Darboux  and E. Goursat as well as in many
recent papers.

An adequate description of the general classification problem for the
MAE is achieved by computing of a complete system of differential
invariants with a complete set of relations between them. Scalar
differential invariants of MAE are interpreted as smooth
functions on a ``diffiety'' quotient $J^\infty(MAE)/Ct$, the object
of the category of differential equations corresponding to MAE (see
\cite{KLV}). This quotient is quite singular and admit a
stratification (like a spaces of orbits, total spaces of general
foliations, etc.). Locally the quotient $J^\infty(MAE)/Ct$ has
the form $\mathcal{E}_\infty$ of the infinite prolongation of a system
of differential equations $\mathcal{E}$, which is defined by the relation
between differential invariants. The stratification of
$\mathcal{E}_\infty$ is given by the singular loci 
$$
\mathcal{E}_\infty \supset
\mathcal{E}_\infty'=sing(\mathcal{E}_\infty)\supset
\mathcal{E}_\infty''=sing(\mathcal{E}_\infty')\supset \ldots
$$
which correspond to a reduction of the number of variables.
Due to a generalized ``Kerr theorem'' (\cite{1}, ch. 7.2.3), this
reduction is provided by a sufficiently large local symmetry group, so
the most ``symmetric'' MAE should correspond to a very ``singular
strata''. We will call such equations ``special MAE''. We will restrict
ourself throughout this paper to the classification problem of these
special MAE.

A modern geometric approach to special MAE was proposed by
V. Lychagin \cite{6} (after an idea of E. Cartan) and was applied to
this classification problem by V. Lychagin and V. Roubtsov in the
papers \cite{1},\cite{7}, and \cite{8}. From the general point of
view, the Monge-Amp\`ere equations should be distinguished by the
differential invariants of order at most $2$, but the approach of
V. Lychagin permits, using the existence of the canonical contact
structure on the space $J^1$, to take into consi\-de\-ration only 
scalar invariants of contact (under some mild restrictions, symplectic)
diffeomorphisms of $J^1$ (cotangent space). 

V.L and V.R established in \cite{1}  an almost complete classification
of special MAE in the dimensions 2 and 3 and a partial classification in higher
dimensions. In the dimension $2$, the equivalence problem of two (generic)
MAE can be reduced to the equivalence problem of two generic structures
(more exactly, to the equivalence of two $G$-structures). The
corresponding scalar differential invariant is given by the
Pfaffian. The integrability is established in \cite{7}  by the
annihilation of the Nijenhuis tensor of
an almost complex structure (for an elliptic MAE) or of an almost
product structure (for a hyperbolic MAE). In the dimension $3$, the
corresponding scalar invariant was proposed in \cite{1} and was
identified recently with the Hitchin functional on the space of
exterior $3$-forms (see \cite{12},\cite{13}). In our paper in progress
(\cite{13}), we adapt the approach to
the equivalence problem of MAE from the viewpoint of equivalence
problem of some geometric structures in dimension $3$ using the
Hitchin-like invariant. The main
goal of the present paper is to complete the list of normal forms for the MAE
in the dimension 3 with respect to the action of the symplectic group
$Sp(6)$.  

Within the classification of 3-dimensional Monge-Amp\`ere
equations  (as it was observed in \cite{1}) a problem of
Geometric Invariant Theory arises: to find the  list of the normal forms  of
effective (primitive) trivectors on $\mathbb{R}^6$ with respect to the
symplectic group. The corresponding complex classification was
obtained by J.I Igusa in \cite{10} and V. Popov in \cite{11} but it
does not help much in the real case we are interested in.

A distinguished MAE in dimension 3 became recently
a subject of considerable interest after the famous paper of R. Harvey and H.B
Lawson  on calibrated geometries (\cite{9}). This is the equation
$$
\hess(h)-\Delta h = 0
$$
(where $\hess(h)$ is the hessian of $h$ and $\Delta$ is the Laplace
operator), which is a criteria for the graph of $\nabla h$ to be
special lagrangian. The normal form associated with this equation was missed in
\cite{1} and motivates the present paper. 

Here we complete the symplectic classification of the effective 3-forms with
constant coefficients and study the special lagrangian
orbit. We calculate its symplectic invariant, its stabilizer and its
Cartan's prolongation. Finally we prove, using the  arguments similar to
\cite{1}, a classification theorem for $3$-dimensional MAE.

This paper has the following structure:

In the first section, we recall Lychagin's approach  to MAE  based on
the differential forms on the $1$-jet space. We also formulate here the
classification problem in an appropriate form. 

The second section  deals with the symplectic classification of
effective $3$-forms on $\mathbb{R}^6$. This classification is obtained
by means of a recursive formula for $3$-forms, which allows us to reduce the
problem to $2$-forms on $\mathbb{R}^4$. So, this  is a little 
bit technical. The main result of it is the theorem $\ref{Classifi}$.

The stabilizers of the different orbits and their Cartan's
prolongations are computed in the third section. We explicitly identify  these
stabilizers with Lie subalgebras of $Sp(6)$ and we summarize
the results in table $\ref{table2}$.

These results are used in the last section  to establish a
local symplectic classification of special MAE. We adapt
here V.L and V.R's results (\cite{1}) to our case which is less
general.

This article constitutes  a part of the author's PhD thesis being  prepared  at 
Angers University. I would like to thank my advisor V. Roubtsov for
suggesting the problem  and helpful discussions. I would like also
to thank professors V.V. Lychagin, A.M. Vinogradov  and I. Roulstone
for their valuable remarks.

\section{Formulation of the problem}

Let $(V,\Omega)$ be a symplectic $2n$-dimensional vector space over
$\mathbb{R}$. Denote by $\Gamma:V\rightarrow V^*$ the isomorphism
determined by $\Omega$. Let $X_\Omega\in\Lambda^2(V)$ be the unique
bivector which satisfies $\Gamma^2(X_\Omega)= \Omega$, where $\Gamma^2:
\Lambda^2(V)\rightarrow \Lambda^2(V^*)$ is the exterior power of
$\Gamma$ \footnote{We denote by $\Lambda^*(V^*)$ the space of exterior
  forms on a vector space $V$ and by $\Omega^*(X)$ the space
  of differential forms on a manifold $X$}.

One can introduce the operators  $\bot:\Lambda^k(V^*)\rightarrow
\Lambda^{k-2}(V^*)$, $\omega\mapsto i_{X_\Omega}\omega$ and $\top: \Lambda^k(V^*)\rightarrow
\Lambda^{k+2}(V^*)$, $\omega\mapsto \omega\wedge \Omega$ (see
\cite{6}). They have the following properties:
$$
\begin{cases}
[\bot,\top](\omega)=(n-k)\omega \;\text{, $\forall \omega\in \Lambda^k(V^*)$};&\\
\bot: \Lambda^k(V^*)\rightarrow \Lambda^{k-2}(V^*)\text{ is into
  for } k\geq n+1;&\\
\top: \Lambda^k(V^*)\rightarrow \Lambda^{k+2}(V^*)\text{ is onto
  for } k\leq n-1.&\\
\end{cases}
$$
We will say that a $k$-form $\omega$ is effective if
$\bot\omega=0$ and we will denote by $\Lambda^{k}_\varepsilon(V^*)$ the
vector space of effective $k$-forms on $V$. When  $k=n$, $\omega$ is
effective if and only if $\omega\wedge \Omega=0$.

The next theorem explains the fundamental role played by the effective
forms in the theory of Monge-Amp\`ere operators (see \cite{6}):

\begin{Theo}[Hodge-Lepage-Lychagin]{\label{hodge}}
\begin{enumerate}
\item Every form $\omega\in \Lambda^k(V^*)$ can be uniquely decomposed
  into the finite sum
$$
\omega= \omega_0+\top \omega_1+\top^2\omega_2+\ldots,
$$
where all $\omega_i$ are effective forms.
\item If two  effective $k$-forms vanish on the same $k$-dimensional
  isotropic vector subspaces in $(V,\Omega)$, they are proportional.
\end{enumerate}
\end{Theo}

Let $M$ be an $n$-dimensional smooth manifold. Denote by $J^1M$
the space of $1$-jets of smooth functions on $M$. Let $j^1(f): 
M\rightarrow J^1M,  x\mapsto [f]_{x}^1$ be the natural section
associated with the smooth function $f$ on $M$. The Monge-Amp\`ere operator
$\Delta_\omega: C^\infty(M)\rightarrow \Omega^n(M)$ associated to a
differential $n$-form $\omega\in \Omega^n(J^1M)$ is the differential
operator defined by 
$$
\Delta_\omega(f)=j_1(f)^*(\omega)
$$

Let $U$ be the contact form on $J^1M$ and $X_1$ be the Reeb's vector
field. Denote $C(x) = Ker(U_x)$ for $x\in J^1M$. Note that  $(C(x),dU_x)$
is a $2n$-dimensional symplectic vector space and 
$$
T_xJ^1M= C(x)\oplus \mathbb{R}X_{1x}.
$$
A solution of the equation $\Delta_\omega=0$ is a
legendrian submanifold $L^{2n}$ of $(J^1M,U)$ such that
$\omega|_L=0$. Note that in each point $x\in L$, $T_xL$ is a
lagrangian subspace of $(C(x),dU_x)$. If the projection $\pi:
J^1M\rightarrow M$ is a local diffeomorphism on $L$ then $L$ is
locally the graph of a section $j^1(f)$, where $f$ is a regular
solution of the equation $\Delta_\omega(f)=0$. 

We will denote by $\Omega^*(C^*)$ the space of differential forms vanishing
on $X_1$. In each point $x$, $(\Omega^k(C^*))_x$ can be naturally
identified with $\Lambda^k(C(x)^*)$.  Let $\Omega_\varepsilon^*(C^*)$
be the space of forms which are effective on $(C(x), dU_x)$ in each
point $x\in J^1M$. The first part of the theorem \ref{hodge} means that 
$$
\Omega^*_\varepsilon(C^*)=\Omega^*(J^1M)/I_C,
$$
where $I_C$ is the Cartan ideal generated by $U$ and $dU$. The second
part means that two differential $n$-forms $\omega$ and $\theta$ on
$J^1M$ determine the same Monge-Amp\`ere operator if and only if
$\omega-\theta\in I_C$.

$Ct(M)$, the pseudo-group of contact diffeomorphisms on $J^1M$, naturally acts on
the set of Monge-Amp\`ere operators in the following way
$$
F(\Delta_\omega)=  \Delta_{F^*(\omega)},
$$
and the corresponding infinitesimal action is
$$
X(\Delta_\omega) = \Delta_{L_X(\omega)}.
$$
We are interested in the problem of local classification of
Monge-Amp\`ere operators in the dimension $3$ with respect to the
action of the contact group. More precisely, we are interested in
the symplectic operators, i.e., operators which satisfy
$$
X_1(\Delta_\omega) = \Delta_{L_{X_1}(\omega)}=0.
$$
Let $T^*M$ be the cotangent space and $\Omega$ be the canonical
symplectic form on it. Let us consider the projection $\beta:
J^1M\rightarrow T^*M$, defined by the following commutative diagram:

$$
\xymatrix{
  \mathbb{R}&&J^1M\ar[ll]_\alpha\ar[rr]^\beta&&T^*M\\
&&&&\\
&&M\ar[lluu]^f\ar[uu]_{j^1(f)}\ar[rruu]_{df}&&\\
}
$$

Using $\beta$, we can naturally identify the space $\{\omega\in
\Omega^*_\varepsilon(C^*): L_{X_1}\omega=0\}$ with the space of
effective forms on $(T^*M,\Omega)$.  Therefore, to classify the
different orbits of symplectic Monge-Amp\`ere operators on a smooth
manifold $M^n$ with respect to the contact group, it is sufficient to
classify the different orbits of the effective $n$-forms on the
cotangent space $T^*M$ with respect to the symplectic group.

\section{Symplectic classification  of effective $3$-forms in the dimension $6$}

Let $(V,\Omega)$ be a $2n$-dimensional symplectic vector space over
$\mathbb{R}$. We will denote the vectors of $V$ by block letters and their
images by $\Gamma$ by small letters: if $A\in V$ then $a\in V^*$ is
defined by 
$$
a(B)=\Omega(A,B).
$$
A basis $(A_1,\ldots,A_n,B_1,\ldots,B_n)$ of  $V$
will be called symplectic if
$$
\Omega = a_1\wedge b_1+ \ldots + a_n\wedge b_n.
$$

\subsection{A recursive formula.\\}

Let $\omega\in \Lambda^{n}_\varepsilon(V^{*})$. Choose  two vectors $A$ and
$B$ such that $\Omega(A,B)=1$ and let $W$ be the skew
orthogonal to $\mathbb{R}A\oplus \mathbb{R}B$ with respect to
$\Omega$. Denote by $\Omega'$ the restriction of $\Omega$
on $W$ and by $\top'$ and $\bot'$ the corresponding operators.

The form $\omega$ can be uniquely decomposed in the following way
$$
\omega = \omega_0\wedge a\wedge b+ \omega_1\wedge a+ \omega_2\wedge b+ 
\omega_3,
$$
with $\omega_i\in \Lambda^*(W^*)$. Moreover, $\omega$ is effective,
therefore we obtain:
\begin{enumerate}
\item  $\bot'\omega_0=\bot'\omega_1=\bot'\omega_2=0$
\item $\omega_3 = -\top'\omega_0 = -\omega_0\wedge \Omega'$
\end{enumerate}

\begin{Prop}{\label{decomposition}}
In the symplectic decomposition  $V = W\oplus(\mathbb{R}A\oplus
\mathbb{R}B)$,  $\omega$ can be expressed in a unique way:
$$
\omega = \omega_0\wedge (a\wedge b-\Omega')+ \omega_1\wedge a+
\omega_2\wedge b,
$$
where $\omega_0$, $\omega_1$ and $\omega_2$ are effective on $(W,\Omega')$.
\end{Prop}

From this proposition it follows that
$$
\Lambda^{n}_\varepsilon(\mathbb{R}^{2n})=\Lambda^{n-2}_\varepsilon(\mathbb{R}^{2(n-1)})\oplus\Lambda^{n-1}_\varepsilon(\mathbb{R}^{2(n-1)})\oplus
\Lambda^{n-1}_\varepsilon(\mathbb{R}^{2(n-1)}),
$$
and in particular,
$$
\Lambda^3_\varepsilon(\mathbb{R}^6) = \mathbb{R}^4\oplus 
\Lambda^{2}_\varepsilon(\mathbb{R}^4)\oplus 
\Lambda^{2}_\varepsilon(\mathbb{R}^4).
$$
Consequently, the dimension of $\Lambda^3_\varepsilon(\mathbb{R}^6)$ is
$4+5+5=14$ . It is worth mentioning that the space
$\Lambda^2_\varepsilon(\mathbb{R}^6)$ is also $14$-dimensional, since
it  is the euclidean orthogonal to $\Omega$ in the Lie algebra $so(6)$ with
respect to the canonical scalar product. 

\subsection{Effective $2$-forms in the dimension $4$.\\}

The symplectic classification of the effective $2$-forms in the dimension
$4$ is well-known (see for instance \cite{7}): 

\begin{Prop}{\label{22}}

Let $(V,\Omega)$ be a $4$-dimensional symplectic vector space and
$\omega\in \Lambda_\varepsilon^2(V^*)$. If $\omega\neq 0$ then there
exists a symplectic basis $(A_1,A_2,B_1,B_2)$ of $V$ such that:

\begin{enumerate}

\item $\omega = \mu(a_{1}\wedge a_2 - b_{1}\wedge b_2)$, $\mu\in
  \mathbb{R}^*$, if  $\omega$ is elliptic, i.e., $Pf(\omega)>0$; 
\item  $\omega = \mu(a_{1}\wedge a_2+b_1\wedge b_2)$, $\mu\in
  \mathbb{R}^*$, if  $\omega$ is hyperbolic, i.e., $Pf(\omega)<0$; 
\item $\omega = a_{1}\wedge a_{2}$, if $\omega$ is parabolic,
  i.e., $Pf(\omega)=0$.
\end{enumerate}
Here the pfaffian $Pf(\omega)$  is the symplectic
invariant  defined by 
$$
\omega\wedge \omega=Pf(\omega)\Omega\wedge \Omega.
$$
\end{Prop}

Note that this proposition can be immediately deduced from
$\eqref{decomposition}$.

\subsection{Effective $3$-forms in the dimension $6$.\\}

Now let $\omega\in\Lambda^3_\varepsilon(V^6)$ be an effective
$3$-form in the dimension $6$.

Following V. Lychagin and  V. Roubtsov (\cite{1}), let us associate to
$\omega$ the quadratic form $q_\omega\in S^2(V^*)$:

$$
q_\omega(X)=  -\frac{1}{4}\bot^{2}\omega_{X}^2,
$$
with  $\omega_X = i_X\omega\in \Lambda^2(V^*)$.

In fact, this form gives us the roots of the characteristic
polynom of $\omega$:
$$
P_{\omega_X}(\lambda)=\frac{(\omega_X-\lambda\Omega)^3}{\Omega^3}=\lambda(\lambda-
\sqrt{q_\omega(X)})(\lambda+ \sqrt{q_\omega(X)}). 
$$

Moreover, if  $F\in Sp(6)$ then one has
$$
q_{F^*\omega}(X) = q_\omega(F^{-1}X).
$$

This  invariant provides us with some information about the decomposition
$\ref{decomposition}$ of $\omega$. More precisely, if in the
symplectic decomposition $V=W\oplus (\mathbb{R}A\oplus \mathbb{R}B)$
$$
\omega = \omega_0\wedge (a\wedge b-\Omega')+ \omega_1\wedge a+
\omega_2\wedge b
$$
holds, then 
$$
\begin{cases}
q_\omega(A) = -\frac{1}{4}\bot^2(\omega_2\wedge \omega_2)&\\
q_\omega(A,B)=\frac{1}{4}\bot^2(\omega_1\wedge \omega_2)
\end{cases}
$$
(we denote the symmetric bilinear form associated with
$q_\omega$ also by $q_\omega$).

Therefore, one obtains:
\begin{equation}{\label{caract_inva}}
\begin{cases}
q_\omega(A)=0\Leftrightarrow \text{ $\omega_2$ is degenerate  on
  $W$}&\\
q_\omega(A,B)=0\Leftrightarrow \omega_1\wedge \omega_2 =0&\\  
\end{cases}
\end{equation}

Now let us consider  two distinct cases: $q_\omega=0$ and $q_\omega\neq 0$.

\subsubsection{$q_\omega=0$}

\begin{Lem}{\label{omega_X}}
If  $\omega\in\Lambda^3_\varepsilon(V^6)$ satisfies $\omega\wedge
\omega_X=0$ for all $X\in V$, then there exists  $X\in V-\{0\}$ such
that $\omega_X=0$.
\end{Lem}

\begin{proof}
Let  $A\in V-\{0\}$: $\omega\wedge \omega_A = 0$. This leads to $\omega_A\wedge
\omega_A=0$ and, therefore, $\omega_A$ must have the form   $\omega_A =
\theta_1\wedge \theta_2$. Consequently,  $dim\{B: \omega_{A\wedge
  B}=0\}\geq 4$ and there exists  $B\in V$ such that $\Omega(A,B)=1$ and
$\omega_{A\wedge B}=0$. Therefore, in the symplectic decomposition
$V=W\oplus (\mathbb{R}A\oplus \mathbb{R}B)$ we obtain
$$
\omega = \omega_1\wedge a+ \omega_2\wedge b.
$$
Moreover, $q_\omega(B)=0$, so $\omega_1\wedge \omega_1=0$ and then
$\omega_1=a_1\wedge a_2$ with  $A_1,A_2\in W$. Similarly, $\omega_2 =
b_1\wedge b_2$ with  $B_1,B_2\in W$. Since both $\omega_1$ and $\omega_2$
are effective, $\Omega(A_1,A_2)=\Omega(B_1,B_2)=0$. We can
suppose, for example, that $A_1\neq 0$. If $\omega_{A_1}=0$ then the
proof is finished. If not, since $\omega_{A_1} =
\Omega(B_1,A_1)b_2\wedge b - \Omega(B_2,A_1)b_1\wedge b$, we can
suppose that $\Omega(B_1,A_1)\neq 0$. Denote then: 
$$
B'_2 = B_2-\frac{\Omega(B_2,A_1)}{\Omega(B_1,A_2)}B_1
$$
Since $\Omega(B'_2,A_1)=0$, $\omega_{A_1} = \Omega(B_1,A_1)b'_2\wedge
b$. Then necessarily  $B'_2\neq 0$. Recall that  $\omega\wedge
\omega_{A_1}=0$ so we have $a_1\wedge a_2\wedge b'_2=0$, i.e. $B'_2=\alpha_1
A_1+\alpha_2 A_2$ and then, 
$$
\Omega(B'_2,A_1)=\Omega(B'_2,A_2)=\Omega(B'_2,B_1)=0.
$$
This implies that $\omega_{B'_2}=0$.
\end{proof}

\begin{Prop}{\label{q=0}}
Let $\omega\in\Lambda_{\varepsilon}^3(V^6)$ with $q_\omega=0$ and
$\omega\neq 0$. Then there exists a symplectic basis
$(A_1,A_2,A_3,B_1,B_2,B_3)$ of  $(V,\Omega)$ in which
$$
\omega = a_{1}\wedge a_{2}\wedge a_{3}.
$$
\end{Prop}

\begin{proof}

For all $X$ and  $Y$,
$\bot^2(\omega_X\wedge\omega_Y)=0$. Consequently, $\forall X,Y\in V$,
$\bot^2\top(\omega_X\wedge \omega_Y)=\top\bot^2\omega_X\wedge
\omega_Y=0$. Recall that  $\bot:\Lambda^{3+i}(V^6)\rightarrow
\Lambda^{1+i}(V^6)$ is into. This leads to the relation  
$$
\Omega\wedge \omega_X\wedge \omega_Y=0,
$$
which holds for all $X,Y\in V$. Note that $\Omega\wedge \omega_X=-\Omega_X\wedge
\omega$ since $\omega$ is effective. Therefore, $\Omega_X\wedge \omega\wedge
\omega_Y=0$, for all $X,Y\in V$ and  $\omega\wedge
\omega_Y=0$ holds for all  $Y\in V$.  Applying lemma \ref{omega_X}, we can
deduce that there exists  $A\in V-\{0\}$ such that
$\omega_A=0$. Let then $B\in V$ be such that $\Omega(A,B)=1$. In the
symplectic decomposition $W\oplus(\mathbb{R}A\oplus\mathbb{R}B)$ we obtain 
$$
 \omega = \omega_1\wedge a,
$$
where $\omega_1$ is an effective form on $W$. Moreover,  $\omega_1$
is parabolic on  $W$, since $q_\omega(B)=0$. From
$\ref{22}$ one can conclude that there exists a symplectic basis
$(A_1,A_2,B_1,B_2)$ of  $W$ in which 
$$
\omega_1=a_1\wedge a_2.
$$

\end{proof}

\subsubsection{$q_\omega\neq 0$.\\}

Consider a non-zero vector $A\in V$ and denote $E_A = Ker(\omega_A: V\rightarrow V^*)$.

\begin{Lem}
If $q_\omega(A)\neq 0$ then $dim(E_A)=2$.
\end{Lem}

\begin{proof}
$\omega_A$ is  non-degenerate  on any subspace $Z$ of $V$ such that
$V=Z\oplus E_A$. Therefore, $dim(E_A)$ must be even. Moreover, $\omega_{A}\wedge
\omega_A\in \Lambda^4(Z^*)$ is  not zero, so $dim(E_A)\leq 2$. Since
$A\in E_A$, $dim(E_A)=2$.
\end{proof}

\begin{Lem}
If $q_\omega(A)\neq 0$ then $E_A$ is not isotropic.
\end{Lem}

\begin{proof}
Let $B$ be a vector satisfying $\Omega(A,B)=1$ and let
$\omega=\omega_0\wedge(a\wedge b-\Omega')+\omega_1\wedge a+
\omega_2\wedge b$ be the symplectic decomposition of $\omega$ in
$W\oplus(\mathbb{R}A\oplus\mathbb{R}B)$. If  $C\in E_A$ with $A\wedge
B\wedge C\neq 0$, we can write $C=D+\alpha A+ \beta B$, $D\in
W-\{0\}$. Then, if we assume that $\beta=0$ for such a $C$, it is
straightforward to obtain
$$
0= i_{A\wedge C}\omega= -i_C(\omega_0\wedge \alpha - \omega_2)= -\omega_0(D)\alpha- i_D\omega_2,
$$
and therefore, $i_D\omega_2=0$. However, since $q_\omega(A)\neq 0$, $\omega_2$
is  non-degenerate  on  $W$ and then $D=0$. It is impossible, so we can
conclude that $\Omega(A,C)=\beta\neq 0$. 
\end{proof}
 
Choose then $B_0\in E_A$ such that $\Omega(A,B)=1$ and denote
$B=B_0+tA$, with $t=-\frac{q_\omega(A,B_0)}{q_\omega(A)}$.  In the
symplectic decomposition $W\oplus (\mathbb{R}A\oplus \mathbb{R}B)$ we have
$$
\omega = \omega_1\wedge a+ \omega_2\wedge b,
$$
where $\omega_1$ and $\omega_2$ are effective on  $W$ and  $\omega_1$
is non-degenerate. At last,  $q_\omega(A,B)$ equals to zero and so
does $\omega_1\wedge \omega_2$. All this can be summarized to the following

\begin{Prop}
Let $\omega\in \Lambda^3_\varepsilon(V^6)$ such that $q_\omega\neq
0$. There  exists a symplectic decomposition  
$V=W\oplus(\mathbb{R}A\oplus\mathbb{R}B)$ in which $\omega$ can be written as
$$
\omega = \omega_1\wedge a+\omega_2\wedge b
$$
Moreover, $\omega_1$ and $\omega_2$ are effective on the symplectic
space  $(W,\Omega')$, while $\omega_2$ and $\Omega'$ are effective on
the symplectic space $(W,\omega_1)$.
\end{Prop}

Since $\Omega'$ is non-degenerate, $\Omega'$ is hyperbolic or
elliptic  on  $(V,\omega_1)$. $\omega_2$ can be hyperbolic, elliptic
or parabolic. A careful study of these six cases allows us to obtain
all possible orbits. This study is a little bit long and tiresome,
so we will only report the details of one case, which is most
exemplary \footnote{See (\cite{1}) for the others}: namely, the case
of  elliptic $\omega_2$ and  $\Omega'$. 

If $\omega_2$ is elliptic, then there  exists a symplectic basis 
$(A_1,A_2,B_1,B_2)$ of $(W,\omega_1)$ in which
$$
\omega_2 = \lambda(a_1\wedge b_2-a_2\wedge b_1), \lambda\neq 0.
$$
After observing that $\Omega'\wedge \omega_1=\Omega'\wedge
\omega_2=0$, we can conclude that
$$
\Omega' = pa_1\wedge a_2+qb_1\wedge b_2+r(a_1\wedge b_2+a_2\wedge b_1)+
s(a_1\wedge b_1-a_2\wedge b_2).
$$
Note that $pq+r^2+s^2<0$ since $\Omega'$ is  elliptic. In  particular,
$q$ cannot be equal to zero.   Let $A_t$ and $B_t$ be the
transformations that depend on the real parameter  $t$ 
$$
A_t = \begin{pmatrix}
1&0&0&t\\
0&1&t&0\\
0&0&1&0\\
0&0&0&1\\
\end{pmatrix},
B_t = \begin{pmatrix}
1&0&t&0\\
0&1&0&-t\\
0&0&1&0\\
0&0&0&1\\
\end{pmatrix}.
$$
($A_t$ and $B_t$ are written in the basis $(A_1,A_2,B_1,B_2)$). $A_t$ and $B_t$
fix $\omega_1,\omega_2$ and act on $\Omega$ in the following way:
$$
\begin{cases}
(p,q,r,s)\overset{A_t}{\longrightarrow} (p-qt^2-2st,q,r,s+qt)&\\
(p,q,r,s)\overset{B_t}{\longrightarrow} (p-qt^2+2rt,q,r,r-qt)&\\
\end{cases}
$$
Let us apply the transformation $B_u$ and then the transformation
$A_v$ with  $u=-\frac{r}{q}$ and  $v=-\frac{s}{q}$. In the new basis we will obtain:
$$
\begin{cases}
\omega_1 = a_1\wedge b_1+a_2\wedge b_2,&\\
\omega_2=\lambda(a_1\wedge b_2-a_2\wedge b_1),&\lambda\neq 0,\\
\Omega' = pa_1\wedge a_2+qb_1\wedge b_2,& pq<0.\\
\end{cases}
$$
After applying the next transformation   
$$
F=\begin{pmatrix}
e^t&0&0&0\\
0&e^t&0&0\\
0&0&e^{-t}&0\\
0&0&0&e^{-t}\\
\end{pmatrix},
$$
with $e^{4t}=-\frac{q}{p}>0$, we will get the following expression for
$\Omega'$: 
$$
\Omega' = \mu(a_1\wedge a_2-b_1\wedge b_2)
$$
(note that $F$ does not change $\omega_1$ and $\omega_2$). 
In the symplectic basis  $(A'_1=A,A'_2=\mu A_1,A'_3=\mu
B_1, B'_1=B,B'_2=A_2,B'_3=-B_2)$ of  $(V,\Omega)$ one obtains
$$
\omega = \frac{1}{\mu^2} a'_1\wedge a'_2\wedge a'_3 - a'_1\wedge b'_2\wedge
b'_3 - \frac{\lambda}{\mu}b'_1\wedge a'_2\wedge b'_3 -
\frac{\lambda}{\mu}b'_1\wedge b'_2\wedge a'_3.
$$
In the  symplectic basis $(A_1,A_2,A_3,B_1,B_2,B_3)$, where
$$
\begin{cases}
A_1 = \mu A'_1,& B_1=\frac{1}{\mu}B'_1,\\
A_2 = \mu\nu A'_2,& B_2 = \frac{1}{\mu\nu}B'_2,\\
A_3 =\nu B'_3,& B_3 = -\frac{1}{\nu}A'_3,
\end{cases}
$$
with  $\frac{\mu}{\lambda} = \varepsilon \nu^2$, $\varepsilon=\pm
1$,  we have
$$ 
\omega = b_3\wedge a_1\wedge a_2-b_2\wedge a_1\wedge a_3+\varepsilon
b_1\wedge a_2\wedge a_3 - \varepsilon\nu^2 b_1\wedge b_2\wedge b_3.
$$
 
Similar results concerning other cases lead us to the

\begin{Theo}{\label{Classifi}}
Let $(e_1,e_2,e_3,f_1,f_2,f_3)$ be a symplectic basis of a $6$
dimensional symplectic vector space $V$. Every effective $3$-form on
$V$ is $Sp(6)$-equivalent  to one and only one form of the table
$\ref{table1}$.
\end{Theo}

\begin{table}[!ht]
\begin{tabular}{|c|c|c|}
\hline
$N^o$& Form& Invariant\\
\hline
1.& $e_{1}^*\wedge e_{2}^*\wedge e_{3}^*+ \gamma f_{1}^*\wedge f_{2}^*\wedge f_{3}^*,\gamma \neq 0$&$2q_\omega=\gamma(e_{1}^*f_{1}^*+ e_{2}^*f_{2}^*+ e_{3}^*f_{3}^*)$\\
\hline
2.&$f_{1}^*\wedge e_{2}^*\wedge e_{3}^*+ f_{2}^*\wedge e_{1}^*\wedge e_{3}^*$& $q_\omega=(e_{1}^*)^2-(e_{2}^*)^2+(e_{3}^*)^2$\\
&$+ f_{3}^*\wedge e_{1}^*\wedge e_{2}^*+ \nu^2 f_{1}^*\wedge f_{2}^*\wedge f_{3}^*,\nu\neq 0$&$+\nu^2((f_{1}^*)^2-(f_{2}^*)^2+(f_{3}^*)^2)$\\
\hline
3.&$f_{1}^*\wedge e_{2}^*\wedge e_{3}^*- f_{2}^*\wedge e_{1}^*\wedge e_{3}^*$& $q_\omega=-(e_{1}^*)^2-(e_{2}^*)^2-(e_{3}^*)^2$\\
&$+ f_{3}^*\wedge e_{1}^*\wedge e_{2}^*- \nu^2 f_{1}^*\wedge f_{2}^*\wedge f_{3}^*,\nu\neq 0$&$+\nu^2(-(f_{1}^*)^2-(f_{2}^*)^2-(f_{3}^*)^2)$\\
\hline
4.&$f_{1}^*\wedge e_{2}^*\wedge e_{3}^*+ f_{2}^*\wedge e_{1}^*\wedge e_{3}^*+ f_{3}^*\wedge e_{1}^*\wedge e_{2}^*$& $q_\omega=(e_{1}^*)^2-(e_{2}^*)^2+(e_{3}^*)^2$\\
\hline
5.&$f_{1}^*\wedge e_{2}^*\wedge e_{3}^*- f_{2}^*\wedge e_{1}^*\wedge e_{3}^*+ f_{3}^*\wedge e_{1}^*\wedge e_{2}^*$& $q_\omega=-(e_{1}^*)^2-(e_{2}^*)^2-(e_{3}^*)^2$\\
\hline
6.&$f_{3}^*\wedge e_{1}^*\wedge e_{2}^*+ f_{2}^*\wedge e_{1}^*\wedge e_{3}^*$& $q_\omega=(e_{1}^*)^2$\\
\hline
7.&$f_{3}^*\wedge e_{1}^*\wedge e_{2}^*- f_{2}^*\wedge e_{1}^*\wedge e_{3}^*$& $q_\omega=-(e_{1}^*)^2$\\
\hline
8.&$e_{1}^*\wedge e_{2}^*\wedge e_{3}^*$& $q_\omega=0$\\ 
\hline
9.& $0$&$q_\omega=0$\\ 
\hline
\end{tabular}
\caption{Classification of  the effective $3$-forms in the dimension $6$}\label{table1}
\end{table}

Note that we have abandoned the notation \emph{block letters}
$\overset{\Gamma}{\longrightarrow}$ \emph{small letters}. In what follows, it
is more convenient to work in the dual basis
$(e_1^*,e_2^*,e_3^*,f_1^*,f_2^*,f_3^*)$. There is an obvious
correspondence between these two notations:
$$
\begin{cases}
\Gamma(e_i) = f_{i}^*&\\
\Gamma(f_i)= -e_{i}^*
\end{cases}
$$

\section{The stabilizers of orbits and their prolongation}

\subsection{Stabilizers  of the orbits.\\}

Let us denote $sp(V)=sp(6)$ the Lie algebra of $Sp(V)$ and consider its
action on $\Lambda^*(V^*)$, induced by the action of $Sp(V)$:
$$
X.\omega = L_X\omega
$$
We compute here the stabilizers of the most significant  orbits. Recall
that the stabilizer of a form $\omega$ is:
$$
\mathcal{J}_\omega = \{X\in sp(V):  L_X\omega=0\}.
$$
Our forms have constant coefficients, therefore $L_X\omega =
d(i_X\omega)$. Straightforward computation (made with help of Maple
V) shows the following

\begin{Prop}
The stabilizers of forms $1$ to $5$ listed in  table   $\ref{table1}$ are:
\begin{enumerate}
\item $\omega=e_{1}^*\wedge e_{2}^*\wedge e_{3}^*+\gamma f_{1}^*\wedge f_{2}^*\wedge f_{3}^*$, $\gamma\neq 0$
\begin{displaymath}
\mathcal{J}_\omega=\{\begin{pmatrix}
B&0\\0&-B^t
\end{pmatrix}:  B\in sl(3,\mathbb{R})\};
\end{displaymath}
\item $\omega=f_{1}^*\wedge e_{2}^*\wedge e_{3}^*+f_{2}^*\wedge e_{1}^*\wedge e_{3}^*+f_{3}^*\wedge e_{1}^*\wedge e_{2}^*+\nu^2 f_{1}^*\wedge f_{2}^*\wedge f_{3}^*$, $\nu\neq 0$
\begin{displaymath}
\mathcal{J}_\omega=\{\begin{pmatrix}
0&\alpha&\beta&\lambda_1&\xi_1&\xi_2\\ 
\alpha&0&\gamma&\xi_1&\lambda_2&\xi_3\\
-\beta&\gamma&0&\xi_2&\xi_3&\lambda_3\\
-\nu^2\lambda_1&\nu^2\xi_1&-\nu^2\xi_2&0&-\alpha&\beta\\
\nu^2\xi_1&-\nu^2\lambda_2&\nu^2\xi_3&-\alpha&0&-\gamma\\
-\nu^2\xi_2&\nu^2\xi_3&-\nu^2\lambda_3&-\beta&-\gamma&0
\end{pmatrix}:  \lambda_1-\lambda_2+\lambda_3=0\};
\end{displaymath}

\item $\omega=f_{1}^*\wedge e_{2}^*\wedge e_{3}^*-f_{2}^*\wedge e_{1}^*\wedge e_{3}^*+f_{3}^*\wedge e_{1}^*\wedge e_{2}^*-\nu^2 f_{1}^*\wedge f_{2}^*\wedge f_{3}^*$, $\nu\neq 0$
\begin{displaymath}
\mathcal{J}_\omega=\{\begin{pmatrix}
0&\alpha&\beta&\lambda_1&\xi_1&\xi_2\\ 
-\alpha&0&\gamma&\xi_1&\lambda_2&\xi_3\\
-\beta&-\gamma&0&\xi_2&\xi_3&\lambda_3\\
-\nu^2\lambda_1&-\nu^2\xi_1&-\nu^2\xi_2&0&\alpha&\beta\\
-\nu^2\xi_1&-\nu^2\lambda_2&-\nu^2\xi_3&-\alpha&0&\gamma\\
-\nu^2\xi_2&-\nu^2\xi_3&-\nu^2\lambda_3&-\beta&-\gamma&0
\end{pmatrix}:  \lambda_1+\lambda_2+\lambda_3=0\};
\end{displaymath}

\item $\omega=f_{1}^*\wedge e_{2}^*\wedge e_{3}^*+f_{2}^*\wedge e_{1}^*\wedge e_{3}^*+f_{3}^*\wedge e_{1}^*\wedge e_{2}^*$
\begin{displaymath}
\mathcal{J}_\omega=\{\begin{pmatrix}
0&\alpha&\beta&\lambda_1&\xi_1&\xi_2\\ 
\alpha&0&\gamma&\xi_1&\lambda_2&\xi_3\\
-\beta&\gamma&0&\xi_2&\xi_3&\lambda_3\\
0&0&0&0&-\alpha&\beta\\
0&0&0&-\alpha&0&-\gamma\\
0&0&0&-\beta&-\gamma&0
\end{pmatrix}:  \lambda_1-\lambda_2+\lambda_3=0\};
\end{displaymath}

\item $\omega=f_{1}^*\wedge e_{2}^*\wedge e_{3}^*+f_{2}^*\wedge e_{1}^*\wedge e_{3}^*-f_{3}^*\wedge e_{1}^*\wedge e_{2}^*$
\begin{displaymath}
\mathcal{J}_\omega=\{\begin{pmatrix}
0&\alpha&\beta&\lambda_1&\xi_1&\xi_2\\ 
\alpha&0&\gamma&\xi_1&\lambda_2&\xi_3\\
\beta&-\gamma&0&\xi_2&\xi_3&\lambda_3\\
0&0&0&0&-\alpha&-\beta\\
0&0&0&-\alpha&0&\gamma\\
0&0&0&-\beta&-\gamma&0
\end{pmatrix}:  \lambda_1-\lambda_2-\lambda_3=0\};
\end{displaymath}
\end{enumerate}
\end{Prop}

\subsection{Prolongation of stabilizers.\\}

We are interested now in the prolongation of these stabilizers. The
prolongation of a linear subspace $\mathcal{J}$ of  $Hom(V,W)$ is (\cite{3}):
$$
\mathcal{J}^{(1)}=  \{T\in Hom(V,\mathcal{J}):  Tuv=Tvu \;\forall 
u,v\in V\}= (W\otimes S^{2}(V^*))\cap (\mathcal{J}\otimes V^*).
$$

In our case $V=W$ and  $\mathcal{J}=\mathcal{J_\omega}$ is a subspace
of $sp(V)$. An element $\theta \in \mathcal{J}_\omega\otimes V^*\subset
sp(V)\otimes V^*\subset V\otimes V^*\otimes V^*$ can be described as follows:
\begin{displaymath}
\begin{aligned}
\theta& = \sum_{i,j,k=1}^3 b_{ij}^k e_i\otimes e_{j}^*\otimes e_{k}^*-a_{ij}^k e_i\otimes f_{j}^*\otimes e_{k}^*+c_{ij}^k f_i\otimes e_{j}^*\otimes e_{k}^*-b_{ji}^k f_i\otimes f_{j}^*\otimes e_{k}^*\\
&+b_{ij}^{k+3} e_i\otimes e_{j}^*\otimes f_{k}^*-a_{ij}^{k+3}
e_i\otimes f_{j}^*\otimes f_{k}^*+c_{ij}^{k+3} f_i\otimes e_{j}^*\otimes f_{k}^*-b_{ji}^{k+3} f_i\otimes f_{j}^*\otimes f_{k}^*\\,
\end{aligned}
\end{displaymath}
with $\begin{pmatrix} B_k&-A_k\\C_k&-B^{t}_k\end{pmatrix}\in
\mathcal{J}_\omega\subset sp(V)$ for $k=1\ldots 6$.
Note that  $\theta\in \mathcal{J}_{\omega}^{(1)}$ if and only if

$$
\begin{cases}
\theta(e_j,e_k)=\theta(e_k,e_j),&\\
\theta(f_j,f_k)=\theta(f_k,f_j),&\\
\theta(e_j,f_k)=\theta(f_k,e_j).&\\
\end{cases}
$$ 

If we take into account the next four relations
\begin{displaymath}
\begin{cases}
\theta(e_j,e_k)=\overset{3}{\underset{i=1}{\sum}} b_{ij}^k e_i+ c_{ij}^k f_i,&\\
\theta(f_j,f_k)=-\overset{3}{\underset{i=1}{\sum}} a_{ij}^{k+3} e_i+ b_{ji}^{k+3} f_i,&\\
\theta(e_j,f_k)=\overset{3}{\underset{i=1}{\sum}} b_{ij}^{k+3} e_i+ c_{ij}^{k+3} f_i,&\\
\theta(f_k,e_j)=-\overset{3}{\underset{i=1}{\sum}} a_{ik}^j e_i+ b_{ki}^j f_i,&\\
\end{cases}
\end{displaymath}
we can conclude  that $\theta\in \mathcal{J}^{(1)}_\omega$ if and only if for all
$i,j,k=1\ldots 3$ the equalities
$$
\begin{aligned}
b_{ij}^k&=b_{ik}^j\label{a},\\
c_{ij}^k&=c_{ik}^j\label{b},\\
a_{ij}^{k+3}&=a_{ik}^{j+3}\label{c},\\
b_{ji}^{k+3}&=b_{ki}^{j+3}\label{d},\\
b_{ij}^{k+3}&=-a_{ik}^j\label{e},\\
b_{ki}^j&=-c_{ij}^{k+3}\label{f},
\end{aligned}
$$
are satisfied.
This allows us to check that  $\mathcal{J}^{(1)}_\omega= 0$ for the
five first  forms $\omega_1,\ldots,\omega_5$ listed in the table \ref{table1}. 
These results are summed up in the table \ref{table2}. 

\begin{table}[!ht]
{\footnotesize
\begin{tabular}{|c|c|c|c|}
\hline
$N^o$&$\mathcal{J}_\omega$&Generators&$\mathcal{J}_{\omega}^{(1)}$\\
\hline
1.&$sl(3,\mathbb{R})$& $\overset{3}{\underset{i,j=1}{\sum}} b_{ij}e_{i}^*f_{j}^* $, $b_{11}+b_{22}+b_{33}=0$&$0$\\
\hline
&&$\alpha(e_{1}^*f_{2}^*+e_{2}^*f_{1}^*)+\beta(e_{1}^*f_{3}^*-e_{3}^*f_{1}^*)+\gamma(e_{2}^*f_{3}^*+e_{3}^*f_{2}^*)$&\\
2.&$su(2,1)$&$+\overset{3}{\underset{i=1}{\sum}}\lambda_i(e_{i}^{*2}+\nu^2f_{i}^{*2})+\xi_1(e_{1}^*e_{2}^*-\nu^2f_{1}^*f_{2}^*)+\xi_2(e_{1}^*e_{3}^*+\nu^2f_{1}^*f_{3}^*)$&$0$\\
&&$+ \xi_3(e_{2}^*e_{3}^*-\nu^2f_{2}^*f_{3}^*)$, $\lambda_1-\lambda_2+\lambda_3=0$&\\
\hline
&&$\alpha(e_{1}^*f_{2}^*-e_{2}^*f_{1}^*)+\beta(e_{1}^*f_{3}^*-e_{3}^*f_{1}^*)+\gamma(e_{2}^*f_{3}^*-e_{3}^*f_{2}^*)$&\\
3.&$su(3)$&$+\overset{3}{\underset{i=1}{\sum}}\lambda_i(e_{i}^{*2}+\nu^2f_{i}^{*2})+\xi_1(e_{1}^*e_{2}^*+\nu^2f_{1}^*f_{2}^*)+\xi_2(e_{1}^*e_{3}^*+\nu^2f_{1}^*f_{3}^*)$&$0$\\
&&$+ \xi_3(e_{2}^*e_{3}^*+\nu^2f_{2}^*f_{3}^*)$, $\lambda_1+\lambda_2+\lambda_3=0$&\\
\hline
&&$\alpha(e_{1}^*f_{2}^*+e_{2}^*f_{1}^*)+\beta(e_{1}^*f_{3}^*-e_{3}^*f_{1}^*)+\gamma(e_{2}^*f_{3}^*+e_{3}^*f_{2}^*)$&\\
4.&$H_2(2,1)\rtimes_\alpha
  so(2,1)$&$+\overset{3}{\underset{i=1}{\sum}}\lambda_i(e_{i}^{*2}+\nu^2f_{i}^{*2})+\xi_1e_{1}^*e_{2}^*+\xi_2e_{1}^*e_{3}^*$&$0$\\
&&$+ \xi_3e_{2}^*e_{3}^*$, $\lambda_1-\lambda_2+\lambda_3=0$&\\
\hline
&&$\alpha(e_{1}^*f_{2}^*+e_{2}^*f_{1}^*)+\beta(e_{1}^*f_{3}^*+e_{3}^*f_{1}^*)+\gamma(e_{2}^*f_{3}^*-e_{3}^*f_{2}^*)$&\\
5.&$H_2(1,2)\rtimes_\alpha
  so(1,2)$&$+\overset{3}{\underset{i=1}{\sum}}\lambda_i(e_{i}^{*2}+\nu^2f_{i}^{*2})+\xi_1e_{1}^*e_{2}^*+\xi_2e_{1}^*e_{3}^*$&$0$\\
&&$+ \xi_3e_{2}^*e_{3}^*$, $\lambda_1-\lambda_2-\lambda_3=0$&\\
\hline
\end{tabular}
}
\caption{Stabilizers of the effective  $3$-forms in the dimension $6$ and their prolongation}\label{table2}
\end{table}

In this table we have identified $sp(\Omega)$ with $S^2(V^*)$ by means
of the canonical isomorphism $h\in S^2(V^*) \mapsto X_h\in sp(\Omega)$.

\section{Local symplectic classification of special Monge-Amp\`ere equations}

Now  we are going to establish  the conditions under which an effective
$n$-form $\omega_2\in\Omega_{\varepsilon}^n(T^*M)$ with non-constant
coefficients is in the same orbit as an effective $n$-form
$\omega_1\in\Omega_{\varepsilon}^n(T^*M)$ with constant coefficients.

This problem is equivalent to the resolution of the differential
equation $\Sigma\subset J^1(2n,2n)$ defined by
$$
\Sigma=\{[F]_q^1: [F^*\omega_1-\omega_2]_q^0=0\text{ and } [F^*\Omega-\Omega]_q^0=0\}. 
$$

Recall that $J^1(m,m)$ is the space of $1$-jets of smooth maps
$\mathbb{R}^m\rightarrow \mathbb{R}^m$ with the canonical
system of coordinates $(q,u,p)$:
$$
\begin{cases}
q_i([F]_q^1)=q_i&i=1\ldots m;\\
u_i([F]_q^1)=F_i(q)&i=1\ldots m;\\
p_{ij}([F]_q^1)=\frac{\partial F_i}{\partial q_j}(q)&i,j=1\ldots m;\\
\end{cases}
$$

\subsection{Symbol of $\Sigma$.\\}

To simplify the notations, we put $m=2n$. One can write 
$$
\Sigma =\{L_1=\ldots = L_r=L_{r+1}=\ldots L_{r+s}=0\}\subset J^1(m,m),
$$
with
$$
\begin{cases}
[F^*\omega_1-\omega_2]_q^0=\underset{i=1}{\overset{r}{\sum}} L_i([F]_{q}^1)\alpha_i,&
\{\alpha_i\}_{i=1\ldots r}\text{ basis of } \Lambda^n(\mathbb{R}^m)\\
\lbrack F^*\Omega-\Omega\rbrack_q^0 = \underset{i=1}{\overset{s}{\sum}} L_{r+i}([F]_q^1)\beta_i,&
\{\beta_i\}_{i=1\ldots s} \text{ basis of } \Lambda^2(\mathbb{R}^{m})
\end{cases}
$$

Let $\theta=[F]_{q_0}^1\in \Sigma$. The  symbol $g(\theta)$ of
$\Sigma\in \theta$ is the set of $h=(h_{ij})\in \mathbb{R}^m\otimes
\mathbb{R}^m$, satisfying the relation

\begin{equation}{\label{symbole}}
\sum_{i,j=1}^m h_{ij}\frac{\partial L_k}{\partial p_{ij}}(\theta)=0
\end{equation}
for all  $k=1\ldots r+s$. Let $h\in g(\theta)$. Define
$H:\mathbb{R}^m\rightarrow \mathbb{R}^m$ by
$$
H(q)=(\sum_{j=1}^m h_{1j}(q_j-q_{j}^0),\ldots, \sum_{j=1}^m
h_{mj}(q_j-q_{j}^0))
$$
and put $\phi_t([G]_q^1) = [G+tH]_{q}^1$, $\forall t\in\mathbb{R}$. From
$\eqref{symbole}$ one can deduce  the relations  
$$
\frac{d}{dt} L_k\circ \phi_t(\theta)|_{t=0}=0,
$$
which hold for $k=1\ldots r$. Taking into account that $F_t= F+tH$
and $q_1=F(q_0)$, we obtain
$$
\begin{cases}
\underset{t\rightarrow 0}{\lim} \frac{ (T_{q_0}F_t)^* \omega_{1,q_1} -
  \omega_{2,q_0}}{t}=0&\\
\underset{t\rightarrow 0}{\lim} \frac{ (T_{q_0}F_t)^* \Omega_{q_1} -
  \Omega_{q_0}}{t}=0&\\ 
\end{cases}
$$
In other words,  
$$
\begin{cases}
\underset{t\rightarrow 0}{\lim} \frac{\psi_{t}^*\omega_{1} -
  \psi_{0}^*\omega_1}{t}=0&\\
\underset{t\rightarrow 0}{\lim} \frac{ \psi_t^* \Omega -
  \psi_{0}^*\Omega}{t}=0&\\ 
\end{cases}
$$
where the linear map $\psi_t:\mathbb{R}^m\rightarrow \mathbb{R}^m$ is defined by
$$
\psi_{ij}^t = \frac{\partial F_i}{\partial q_j}(q_0)+ th_{ij}.
$$
Therefore, $L_X\omega_1=L_X\Omega=0$ with $X=(h_{ij})$. This proves
the following

\begin{Prop}
For all $\theta\in \Sigma$ the symbol of $\Sigma\in \theta$ can be
naturally identified with the stabilizer of $\omega_1$:
$$
g(\theta)= \mathcal{J}_{\omega_1}
$$
\end{Prop}

\begin{Rem}
Note that the prolongation of the symbol coincides with the prolongation of the
Lie subalgebra $\mathcal{J}_{\omega_1}$:
$$
g^{(k)}(\theta) = \mathcal{J}^{(k)}_{\omega_1}\text{, $\forall k\in \mathbb{N}$.}
$$
\end{Rem}

\subsection{The bundle $\Sigma^{(k)}\rightarrow \Sigma^{(k-1)}$.\\}

Let us find the obstructions  for the map  $\pi_{k+1,k}:\Sigma^{(k)}\rightarrow
\Sigma^{(k-1)}$ to be surjective. In our case $\Sigma^{(k)}$ is the differential
equation
$$
\Sigma^{(k)} = \{[F]^{k+1}_q: [F^*\omega_1-\omega_2]_{q}^k=0\text{ and
  } [F^*\Omega - \Omega]_{q}^{k} = 0\}.
$$

Let $\theta=[F]_{q_0}^{k}\in\Sigma^{(k-1)}$. If we introduce the notation
$$
\sigma_k(F)=[\omega_1-F^{-1*}\omega_2]_{q_1}^k
$$
and 
$$
\sigma_k(\theta)= \sigma_k(F)\text{ modulo } Im(c_k),
$$
where $c_k: S^{k+2}(\mathbb{R}^m)\rightarrow \Omega^n(\mathbb{R}^{m})$
 is defined by  
$$
c_k(h)=L_{X_h}(\omega_1),
$$
we can formulate the next

\begin{Prop}
Let  $\theta=[F]_{q_0}^k\in\Sigma^{(k-1)}$. The intersection $\pi_{k+1,k}^{-1}(\theta)\cap
\Sigma^{(k)}$ is not empty if and only if $\sigma_k(\theta)=0$.
 \end{Prop}

\begin{proof}
We can assume that $F$ is a polynomial map of degree less than $k$. Denote
$q_1=F(q_0)$. We assume first that there exists
$\theta'=[G]_{q_0}^{k+1}\in \Sigma^{(k)}$ such that
$[G]_{q_0}^k=[F]_{q_0}^k$. Since $(T_{q_0}G)^*\Omega=\Omega$,
$T_{q_0}G$ is an isomorphism and then  $G:
(\mathbb{R}^m,q_0)\rightarrow (\mathbb{R}^{m},q_1)$ is a local
diffeomorphism. Let $\eta$ be defined by
$$
\eta = F\circ G^{-1}.
$$
Denote $P  = [\eta]_{q_1}^{k+1} - id$. $P$  is a  homogeneous polynom
of degree $k+1$. It is easy to check that for any form $\omega$ we have
\begin{equation}{\label{Annexe1}}
[\eta^*\omega]_{q_1}^k = [\omega]_{q_1}^k+ L_X[\omega]_{q_1}^0,
\end{equation}
where  $X = \underset{i=1}{\overset{m}{\sum}}
P_i\frac{\partial}{\partial q_i}$.
Then, one gets
$$
\begin{aligned}
0& = [\omega_1 - G^{-1*}\omega_2]_{q_1}^k= [\omega_1-\eta^*\omega_1]_{q_1}^k+
[\eta^*(\omega_1-F^{-1*}\omega_2)]_{q_1}^k=\\
& = -L_X(\omega_1)+ [\omega_1 - F^{-1 *}\omega_2]_{q_1}^k+
L_X([\omega_1-F^{-1*}\omega_2]_{q_1}^0).
\end{aligned}
$$
However, $[F]_{q_0}^k\in \Sigma^{(k-1)}$, so
$[\omega_1-F^{-1*}\omega_2]_{q_1}^0$ must be zero. This leads to
$$
\sigma_k(F)= L_X\omega_1.
$$
Moreover, we have:
$$ 
L_X\Omega= [\eta^*\Omega-\Omega]_{q_1}^k= [G^{-1*}F^*\Omega - \Omega]_{q_1}^k= [[G^{-1}]_{q_1}^{k+1 *} [F^*\Omega]_{q_0}^k]_{q_1}^k - \Omega.
$$
Recall that $F$ is a polynomial map with $deg(F)\leq k$. Therefore,
$$
[F^*\Omega]_{q_0}^k = [[F]_{q_0}^{k+1 *} [\Omega]_{q_1}^k]_{q_0}^k
=[[F]_{q_0}^{k *} \Omega]_{q_0}^k = [F^*\Omega]_{q_0}^{k-1} = \Omega.
$$
Since $[G]_{q_0}^{k+1}\in \Sigma^{(k)}$, it is easy to check that
$$
L_X\Omega = [G^{-1 *}\Omega]_{q_1}^k = 0.
$$
$X$ is a hamiltonian vector field with the homogeneous coefficents of
degree $k+1$. Consequently, there exists $h\in S^{k+2}(\mathbb{R}^m)$
for which $X=X_h$. 

Conversely,  let us assume that there exists $X=X_h$ with  $h$ in
$S^{k+2}(\mathbb{R}^m)$ such that $\sigma_k(F)=L_X\omega_1$. Put
$\eta=id+P$ with $X = \underset{i=1}{\overset{m}{\sum}}
  P_i\frac{\partial}{\partial q_i}$ and $G = \eta^{-1}\circ
  F$. Similar considerations lead to 
$$
\begin{cases}
\lbrack \omega_1-G^{-1*}\omega_2\rbrack _{q_1}^k = -L_X(\omega_1)+ \sigma_k(F)=0&\\
\lbrack\Omega-G^{-1*}\Omega\rbrack_{q_1}^k = -L_X\Omega = 0.
\end{cases}
$$
Consequently, one can conclude that $[G]_{q_0}^{k+1}\in \Sigma^{(k)}$
and $[G]_{q_0}^{k}=[F]_{q_0}^k$. 
\end{proof}

\begin{Cor}{\label{diffeo}}
If  $\mathcal{J}^{(k)}_{\omega_1}=0$ and $\sigma_k(\theta)=0$ then $\pi_{k+1,k}:
\Sigma^{(k)}\rightarrow \Sigma^{(k-1)}$ is a local diffeomorphism in
a neighbourhood of $\theta_0\in \Sigma^{(k-1)}$. 
\end{Cor}

\begin{proof}
$\pi_{k+1,k}$ is surjective. Moreover,
$$
Ker(T_{\theta_0}\pi_{k+1,k}) =
g^{(k)}(\theta)=\mathcal{J}_{\omega_1}^{(k)}=0.
$$
Therefore, $\pi_{k+1,k}$ is a local diffeomorphism.
\end{proof}

\subsection{Integrability of  $\Sigma$.\\}

Let $\omega_1,\omega_2\in \Omega^3_\varepsilon(\mathbb{R}^6)$
and suppose that $\omega_1$ has constant coefficients and satisfies
the equation $\mathcal{J}_{\omega_1}^{(1)}=0$. Besides, suppose that:
\begin{enumerate}
\item for all $q$ in a neighbourhood of $q_0$, $\omega_2(q)$ is in the
  same orbit as $\omega_1$.
\item for all $\theta$ in a neighbourhood of $\theta_0=[F]_{q_0}^1\in
  \Sigma$, $\sigma_1(\theta)=0$.
\item for all  $\theta'$ in a neighbourhood of $\theta'_0=[F]_{q_0}^2\in \Sigma^{(1)}$, $\sigma_2(\theta)=0$.
\end{enumerate}

It follows  from  $\ref{diffeo}$ that $\pi_{3,2}:
(\Sigma^{(2)},\theta_0'')\rightarrow (\Sigma^{(1)},\theta'_0)$ and $\pi_{2,1}:
(\Sigma^{(1)},\theta_0')\rightarrow (\Sigma,\theta_0)$ are local
diffeomorphisms.
 
Let $D: \theta\mapsto C(\theta)\cap  T_\theta\Sigma^{(1)}$ be the
restriction of the Cartan distribution on  $\Sigma^{(1)}$. Recall that
$C(\theta)$ is the vector space generated by the $T_\theta j_2 G$ where
$[G]_{q}^2=\theta$.  Let  $\theta=[F]_{q}^2\in \Sigma^{(1)}$ be in a
neighbourhood of $\theta'_0$. Choose $F$ such that  $[F]_q^3\in
\Sigma^{(2)}$: $T_\theta j_2 F\subset C(\theta)$. Moreover, $[F]_q^3\in
\Sigma^{(2)}$ if and only if $T_\theta j_2 F\subset
T_\theta\Sigma^{(1)}$. Therefore, 
$$
T_\theta j_2 F\subset D(\theta).
$$
Let $G$ be a map satisfying $[G]_q^2=\theta$. For any  $X\in T_\theta j_2
G\cap T_\theta\Sigma^{(1)}$ there exists $X_F\in T_\theta j_2 F$ such
that $X-X_F\in Ker(T_\theta \pi_{2,1})$. However, since $X,X_F\in
T_\theta\Sigma^{(1)}$, we obtain     
$$
X-X_F\in T_\theta\Sigma^{(1)}\cap
Ker(T_\theta\pi_{2,1})=g^{(1)}(\theta)=\mathcal{J}_{\omega_1}^{(1)}=0,
$$
and then $X\in T_\theta j_2 F$.

Finally,  for all $\theta$ in a neighbourhood of $\theta'_0$ there
exists $F$ such that $D(\theta)=T_\theta j_2(F)$. Therefore, $D$ is completely
integrable on $J^2(6,6)$ and, according to  the Frobenius
theorem, it is completely integrable on $\Sigma^{(1)}$.  

Consequently, there exists a submanifold $L_0\subset \Sigma^{(1)}$ which is an
integral submanifold of $D$  containing $\theta'_0$.  
Locally $L_0=j_2 G$ if and only if $\pi_{2,0}: L_0\rightarrow
\mathbb{R}^6$ is a local diffeomorphism.  However, there exists $F_0$ such
that $T_{\theta'_0}L_0=D(\theta'_0)= T_{\theta'_0} j_2 F_0$ and
$T_{\theta'_0}\pi_{2,0}: T_{\theta'_0}L_0\rightarrow \mathbb{R}^6$ is
an isomorphism: $L_0=j_2 G$ locally.

\begin{Prop}
Let $\omega_2\in\Omega^{3}_\varepsilon(\mathbb{R}^6)$ be a form with
the following local properties:
\begin{enumerate}
\item for every $q$, $\omega_2(q)\in
  \Lambda^3_\varepsilon(\mathbb{R}^6)$ belongs to the $Sp(6)$-orbit of
  a form $\omega_1$ with constant coefficients, satisfying
  $\mathcal{J}^{(1)}_{\omega_1}=0$; 
\item $\sigma_1=\sigma_2=0$.
\end{enumerate}
Then, locally, $\omega_2$ belongs to the orbit of $\omega_1$.
\end{Prop}

\subsection{Another expression for $\sigma_1$ and $\sigma_2$.\\}

Let $\theta=[F]_{q_0}^1\in \Sigma$. We assume that
$F:(\mathbb{R}^6,q_0)\rightarrow (\mathbb{R}^6,q_1)$ is affine. Denote
$$
\lbrack\omega_2\rbrack_{q_0}^1=\omega_2^0+\omega_2^1,
$$ 
where $\omega_2^0=\omega_2(q_0)$ has constant coefficients and
$\omega_2 ^1=[\omega_{2}]^1-\omega_2(q_0)$ has linear
coefficients. 

\begin{Lem}
$\sigma_1(\theta)=0$ if and only if there exists $h\in
S^3(\mathbb{R}^6)$ such that
$$
\omega_{2}^1=L_{X_h} \omega_2^0
$$
\end{Lem}

\begin{proof}

Since $\theta\in \Sigma$, $F$ satisfies: 
$$
\begin{cases}
F^*\omega_1=\omega_{2}^0&\\
F^*\Omega=\Omega
\end{cases}
$$
Therefore,
$$
\sigma_1(F)=[\omega_1-F^{-1 *}\omega_2]_{q_1}^1= \omega_1 -
[[F^{-1}]_{q_1}^{2 *}[\omega_2]_{q_0}^1]_{q_1}^1= \omega_1 - F^{-1
  *}(\omega_2^0+\omega_2^1)= -F^{-1 *} \omega_{2}^1 
$$
and
$$
\omega_2^1=-F^*(\sigma_1(F)).
$$
After observing that for any form $\omega$ the relation
$$
F^*L_Y\omega = L_XF^*\omega
$$
must hold ($TF(X)=Y\circ F$), we get the result.
\end{proof}

Now we can study  $\sigma_2$. Let $\theta\in \Sigma^{(1)}$. Let us
choose a linear map $F:(\mathbb{R}^6,q_0)\rightarrow
(\mathbb{R}^6,q_1)$ for which $[F]_{q_0}^1=\theta$, and consider the
map $G$ such that $[G]_{q_0}^2=\theta$
and $G^{-1}$ is  polynomial of degree less than $2$:
$$
G^{-1}= \eta\circ F^{-1},
$$
 where  $\eta=id +Q$, $Q$ is a homogeneous polynom of degree
 $2$. Let us denote 
$$
V_i(q) = \frac{1}{6}\sum_{j,k,l=1}^6 a_{jkl}^i(q_j-q_j^0)(q_k-q_k^0)(q_l-q_l^0)
$$
for $i=1\ldots 6$, with
$$
a_{jkl}^i=\sum_{m=1}^6 \frac{\partial^2
  Q_i}{\partial q_m\partial q_k}\frac{\partial^2
  Q_m}{\partial q_j\partial q_l}+ \frac{\partial^2
  Q_i}{\partial q_m\partial q_j}\frac{\partial^2
  Q_m}{\partial q_k\partial q_l}+\frac{\partial^2
  Q_i}{\partial q_m\partial q_l}\frac{\partial^2
  Q_m}{\partial q_j\partial q_k}
$$  
and put $U=\underset{i=1}{\overset{6}{\sum}} Q_i\frac{\partial}{\partial
  q_i}$, $V=\underset{i=1}{\overset{6}{\sum}} V_i\frac{\partial}{\partial
  q_i}$.  It is not difficult to check that for any form $\omega_2$
\begin{equation}{\label{Annexe2}}
[\eta^*\omega_2]_{q_0}^2- [\eta^*\omega_2]_{q_0}^1 =
\omega_{2}^2+L_U\omega_{2}^1+\frac{1}{2}(L_UL_U\omega_{2}^0-
L_V\omega_{2}^0).
\end{equation}

Since $[G]_{q_0}^2\in\Sigma^{(1)}$, $[\omega^1-G^{-1
  *}\omega_2]_{q_0}^1=[\Omega- G^{-1 *} \Omega]=0$, i.e., according to
$\eqref{Annexe1}$: 
$$
0=\omega_1-F^{-1 *}([\eta^*\omega_2]_{q_0}^1)= \omega_1-F^{-1
  *}([\omega]_{q_0}^1+ L_U\omega_{2}^0)= -F^{-1}
*(\omega_{2}^1+L_U\omega_{2}^0) 
$$
and then $\omega_{2}^1=-L_U\omega_{2}^0$. In a similar way we can check
that $L_U\Omega=0$: there exists $h\in S^3(\mathbb{R}^6)$ such that $U=X_h$. Moreover, $deg(G^{-1})\leq 2$, so
$$
\lbrack \Omega - G^{-1 *}\Omega]_{q_1}^2= \Omega - [[G^{-1}]^{3
  *}_{q_1} [\Omega]_{q_0}^2]_{q_1}^2= \Omega - G^{-1 *}\Omega= [\Omega
- G^{-1 *} \Omega]_{q_1}^2= 0. 
$$
Therefore,
$$
0= [\Omega-G^{-1 *}\Omega]_{q_1}^2= \Omega - F^{-1 *}
[\eta^*\Omega]_{q_0}^2= \Omega - F^{-1 *}(\Omega -
\frac{1}{2}(L_UL_U\Omega- L_V\Omega))= -F^{-1 *} L_V\Omega 
$$
and $L_V\Omega=0$, i.e. there exists $k\in S^4(\mathbb{R}^6)$ such
that  $V=X_k$. At last, we have checked that 
$$
\sigma_2(G)=-F^{-1 *}(\omega_{2}^2-\frac{1}{2}
(L_UL_U\omega_{2}^0+L_V\omega_{2}^0)).
$$

\begin{Prop}
$\sigma_1([G]_{q_0}^1)=\sigma_2([G]_{q_0}^2)=0$ if and only if there
exist $h\in S^3$ and $k\in S^4$ for which 
$$
\begin{cases}
\omega_{2}^1=L_{X_h}\omega_{2}^0&\\
\omega_{2}^2 = \frac{1}{2}(L_{X_h}\omega_{2}^1+L_{X_k}\omega_{2}^0)
\end{cases}
$$
\end{Prop}

\begin{proof}

Suppose first  that $\sigma_2([G]_{q_0}^2)=0$. Then there exists $W=X_k$ with
$k\in S^4(\mathbb{R}^6)$ such that $\sigma_2(G)=L_W\omega_1$. So, one has
$$
L_W\omega_0 = -F^{-1 *}(\omega_{2}^2-\frac{1}{2}
(L_UL_U\omega_{2}^0+L_V\omega_{2}^0)).
$$
Therefore,
$$
\omega_{2}^2 = \frac{1}{2}(L_{U_0}\omega_{2}^1+ L_{V_0}\omega_{2}^0)
$$
with $U_0= X_h$, $h\in S^3$, $V_0=X_k$, $k\in S^4$ and 
$L_{U_0}\omega_{2}^0=\omega_{2}^1$.

Conversely, if there exist $h\in S^3$ and $k\in S^4$ such that
$$
\begin{cases}
\omega_{2}^1= L_{U_0}\omega_{2}^0&\\
\omega_{2}^2 = \frac{1}{2}(L_{U_0}\omega_{2}^1+ L_{V_0}\omega_{2}^0)
\end{cases}
$$
with $U_0=X_h$ and  $V_0=X_k$, then $L_U\omega_{2}^0=
-L_{U_0}\omega_{2}^0$. It means that  $U+U_0\in
\mathcal{J}^{(1)}_{\omega_{2}^0}=0$ and, therefore, 
$$
\sigma_2(G) = -F^{-1 *}(\omega_{2}^2-\frac{1}{2}
(L_UL_U\omega_{2}^0+L_V\omega_{2}^0)) = -F^{-1 *}L_\frac{V-V_0}{2}
\omega_{2}^0 = L_W\omega_1
$$
with $W = X_K$, $K\in S^4$.
\end{proof}

\begin{Theo}{\label{Classifieq}}

Consider a Monge-Amp\`ere equation in the dimension $3$,
corresponding to an effective $3$-form $\omega$, such that locally:

\begin{enumerate}
\item for all $q$, $\omega(q)$ belongs to
  one of the five first orbits listed in table $\ref{table1}$
\item for all $q$, the exterior form $[\omega]_{q}^2 = \omega^0+
  \omega^1+ \omega^2$ satisfies 
$$
\begin{cases}
\omega^1=L_{X_h}\omega^0&\\
\omega^2 = \frac{1}{2}(L_{X_h}\omega^1+L_{X_k}\omega^0)
\end{cases}
$$
with  $h\in S^3(\mathbb{R}^6)$ and $k\in S^4(\mathbb{R}^6)$.
\end{enumerate}
Then this differential equation is locally $Sp$-equivalent to one of
the following equations: 
$$
\lambda+\hess(h)=0,     \lambda\neq 0
$$
$$
\frac{\partial^2 h}{\partial q_{1}^2}-\frac{\partial^2
  h}{\partial q_{2}^2}+\frac{\partial^2 h}{\partial
  q_{3}^2}+\nu^2 \hess(h)=0,     \nu\neq 0
$$
$$
\frac{\partial^2 h}{\partial q_{1}^2}+\frac{\partial^2
  h}{\partial q_{2}^2}+\frac{\partial^2 h}{\partial q_{3}^2}-\nu^2
\hess(h)=0,     \nu\neq 0
$$
$$
\frac{\partial^2 h}{\partial q_{1}^2}-\frac{\partial^2
  h}{\partial q_{2}^2}+\frac{\partial^2 h}{\partial
  q_{3}^2}=0
$$
$$
\frac{\partial^2 h}{\partial q_{1}^2}+\frac{\partial^2
  h}{\partial q_{2}^2}+\frac{\partial^2 h}{\partial
  q_{3}^2}=0
$$
where $\hess(h)$ is the hessian of $h$.
\end{Theo}

\end{document}